# A molecular-spin photovoltaic device

Xiangnan Sun[1,2], Saül Vélez[3], Ainhoa Atxabal[3], Amilcar Bedoya-Pinto[4], Subir Parui[3], Xiangwei Zhu[1,2], Roger Llopis[3], Fèlix Casanova[3,5], Luis E. Hueso[3,5]*

[1] Key Laboratory of Nanosystem and Hierarchical Fabrication, CAS Center for Excellence in Nanoscience, National Center for Nanoscience and Technology, Beijing 100190, P. R. China

[2] University of Chinese Academy of Sciences, Beijing 100049, P. R. China

[3] CIC nanoGUNE, 20018 Donostia - San Sebastian, Spain

[4] Max Planck Institute of Microstructure Physics, Halle (Saale), Sachsen Anhalt, Germany

[5] IKERBASQUE, Basque Foundation for Science, 48013 Bilbao, Spain

*Correspondence to: Luis E. Hueso (l.hueso@nanogune.eu)

**Abstract**:

We fabricated a $C_{60}$-based molecular spin photovoltaic device that integrated a photovoltaic response with the spin transport across the molecular layer. The photovoltaic response can be modified under the application of a small magnetic field, with a magnetophotovoltage of up to 5% at room temperature. Device functionalities include a magnetic current inverter and the presence of diverging magnetocurrent at certain illumination levels that could be useful for sensing. Completely spin-polarized currents could be created by balancing the external partially spin polarized injection with the photogenerated carriers.

**One Sentence Summary:**
A novel molecular spin photovoltaic device is conceived by integrating both spin and light degrees of freedom.

Molecular materials, such as $C_{60}$ and $Alq_3$, can preserve the spin polarization of electrical carriers for millisecond-long times (*1, 2*). This property, along with their ability to alter metallic ferromagnetic surface states both in energy and spin polarization (*1-6*), make them attractive for use in spin-transport devices. Spin populations can also play a crucial role in electron-hole pair recombination, which could be exploited in optoelectronic devices such as photovoltaics cells (PVs) and light emitting diodes (LEDs) (*10-14*). For example, a simple dependence of the photocurrent with a large external magnetic field of several Tesla was observed in molecular PV devices with materials such as donor-acceptor blends among others (*11-13*). However, a full interplay between light and spin-polarized currents in molecular devices is still lacking.

We present the design and characterization of a molecular spin photovoltaic (MSP) device that has several distinctive characteristics compared with previously reported inorganic spin photovoltaic devices *(15-17)*,. It shows a switchable resistance and PV effect under low magnetic fields at room temperature. It encodes complex device functionalities; for example, we demonstrate a magnetic current inverter. Finally, the device can generate a completely spin-polarized current under specific light irradiation conditions.

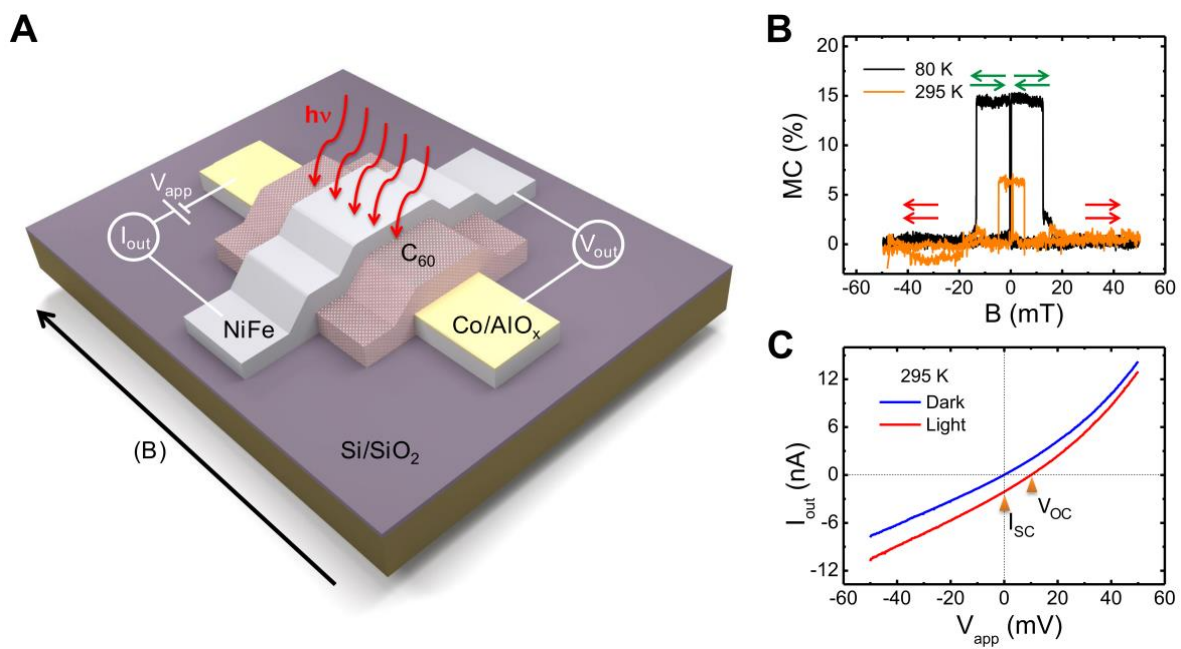

**Fig. 1. Scheme of the device, together with its magnetocurrent and photovoltaic characterization.** (**A**) Schematic representation of the $C_{60}$-based molecular spin photovoltaic device, composed from bottom to top of $Si/SiO_2/Co/AlO_x/C_{60}/Ni_{80}Fe_{20}$. (**B**) Magnetocurrent (*MC*) on the $C_{60}$-based device measured at 295 and 80 K with a bias of 10 mV in dark conditions. The magnetocurrent is defined as the relative change in electrical current as a function of magnetic field, *MC (%) = ($I_P$-$I_{AP}$)/$I_{AP}$×100*; where $I_P$ and $I_{AP}$ represent the current for parallel and antiparallel orientations of the magnetization of the Co and $Ni_{80}Fe_{20}$ electrodes, respectively. The arrows in the image indicate the orientations of the magnetization. (**C**) Current-voltage curves measured with/without white-light irradiation (7.5 mW/cm$^2$) at room temperature when the relative orientation of the electrodes was parallel. The open-circuit voltage ($V_{OC}$) and the short-circuit current ($I_{SC}$) are indicated on the curve. In the measurement setup, the Co electrode is grounded.

The MSP devices have a spin-valve geometry, which is composed of two ferromagnetic metallic (FM) layers (Co and $Ni_{80}Fe_{20}$, respectively) sandwich a $C_{60}$ molecular film, a well-tested material for both photovoltaic (*18-20*) and spintronic applications (*21-24*) [Fig. 1A and figure S1 (*25*)]. We obtained reproducible results for more than ten samples, in part by using a leaky $AlO_x$ barrier (semi-oxidized Al) and a low-temperature molecular growth process (*26,27*) [see a rigid band energy map of the $C_{60}$-based device in Figure S1; device fabrication details are included in the SM (*25*)]. In our devices as in conventional spin valves, one of the FM materials (here Co) injects spin-polarized carriers into the semiconductor layer, and the other FM layer is the spin detector. If the spin polarization of the electrical carriers is preserved across the $C_{60}$ layer, the electrical current flow changes depending on the relative orientation of the magnetization of the FM layers (*23-29*). This change in electric current under the application of a magnetic field can be denoted as magnetocurrent (*MC*) and defined as *MC (%) = ($I_P$-$I_{AP}$)/$I_{AP}$×100*; where $I_P$ and $I_{AP}$ represent the current for parallel and antiparallel orientations of the magnetization respectively.

Our optimized $C_{60}$ devices delivered considerable *MC* values of 6.5% at 295 K and ~ 15% at 80 K (Fig. 1B) [*MC* data at other temperatures are shown in Fig. S2 (*25*)]. These values are in consonance with previous reports for $C_{60}$ layers (*21-24*). Our spin valve structure resembles simple molecular PV cells, and we observed a PV effect under white-light irradiation (7.5 mW/cm$^2$ illuminating an area of 1 cm$^2$) at room temperature (Fig. 1C). The open-circuit voltage ($V_{OC}$) and short-circuit current ($I_{SC}$) of the device, defined as photo-generated bias at zero current and photo-current measured without applied bias respectively, are relatively small at room temperature given the device is a single layer molecular solar cell (*30*), we used low-intensity light irradiation (equivalent to 7.5×10$^{-2}$ Sun), and our ferromagnetic electrodes are not fully transparent [see Figure S3, *25*].

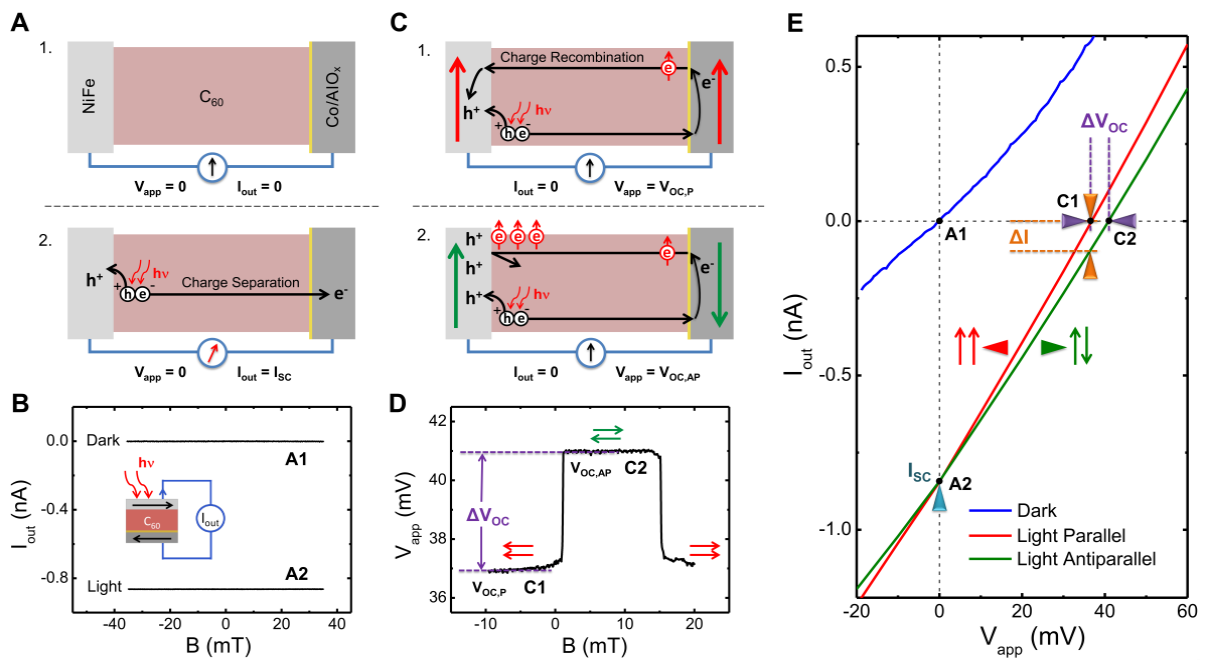

**Figure 2. Device functioning principle, together with its PV characterization in a magnetic field. (A)** Molecular spin photovoltaic (MSP) device operating as a short-circuited single layer molecular solar cell without (Image 1) and with light irradiation (Image 2). **(B)** Current versus magnetic field measurement with/without light irradiation at 80 K (short-circuit

mode). The photo-generated current is not sensitive to the magnetic field. (**C**) MSP device operating in open-circuited mode under the application of a magnetic field. (**D**) Applied voltage ($V_{app}$) versus magnetic field measurement at 80 K (equal to $V_{OC}$ as we operate in open-circuit mode). $V_{OC}$ increased when the magnetic moments of the FM electrodes became antiparallel. (**E**) *I-V* curves measured under an irradiation of white light (7.5 mW/cm$^2$) for the cases in which the magnetic moments of the FM electrodes are parallel and antiparallel. The changes in $V_{OC}$ and $I_{out}$ are indicated on the curves. The measurements shown were performed at 80 K.

We illustrate integrated spin and PV operation in Fig. 2. In a short-circuit solar cell (Fig. 2A), free carriers photogenerated in the C$_{60}$ layer were driven by the built-in potential between the two magnetic electrodes and generated an output current $I_{SC}$ (Fig. 2A-2). These carriers did not produce any spin-mediated magnetoresistance when the magnetization of the FM electrodes was switched with a magnetic field, as they were intrinsically non spin-polarized (see Fig. 2B).

For open-circuit MSP operation, the applied voltage ($V_{app}$) must compensate the photo-generated voltage ($V_{OC}$) as it transports carriers from one magnetic electrode to the opposite while driving charge recombination. For parallel magnetization, the spin-polarized carriers created at the Co electrode could reach the NiFe electrode and recombine with the photo-generated holes (*31, 32*). In this case, $V_{app}$ between the two electrodes corresponded to $V_{OC,P}$ (Fig. 2C-1). For the antiparallel case, the collected holes in the NiFe electrode could only be compensated by the injected spin-polarized electrons in the presence of an enhanced photovoltage ($V_{OC,AP}$) through the spin filtering effect at the FM anode (Fig. 2C-2). The extra generated photovoltage ($\Delta V_{OC}$) is defined as the difference between the photovoltage for antiparallel and parallel configurations of the magnetic electrodes in the MSP device ($\Delta V_{OC} = V_{OC,AP} - V_{OC,P}$). $\Delta V_{OC}$ is the spin-PV response caused by the spin-polarized charge carrier accumulation at the molecule/anode interface (Fig. 2D).

We introduce a magnetophotovoltage (*MPV*) to quantitatively describe the spin-PV effect that represents the photovoltage change under the application of a magnetic field (*MPV* (%) = $\Delta V_{OC}/V_{OC,P} \times 100$). Our MSP device had MPV values of 10.8% at 80 K and 4.6% at room temperature.

All the different effects described above can be experimentally summarized in the *I-V* measurements displayed in Fig. 2E. In the case $V_{app} = 0$, $I_{out}$ was insensitive to the magnetic field and showed a constant value $I_{SC}$ (Fig. 2, A and B). However, for the case of $V_{app} \neq 0$, $I_{out}$ and $V_{OC}$ values were affected by the external magnetic field change (that controlled the parallel/antiparallel alignment of the magnetization of the electrodes), and their change is indicated in the figure as $\Delta I$ and $\Delta V_{OC}$. Because of the increased $V_{OC}$, the MSP device increased in power conversion efficiency (*PCE*) at 80 K from $5.2 \times 10^{-6}$ to $5.9 \times 10^{-6}$ (*33*).

Other functionalities can also be achieved that arise from the relation between spin transport and light irradiation in our devices. In Fig. 3A we show the current versus magnetic field (*I-B*) measurements of the $C_{60}$-based device obtained under a constant 10 mV voltage bias for various intensities of the irradiated light. The output current could be shifted by the PV effect as the irradiated light intensity increased and its sign adjusted to be either negative or positive. At the specific range of light illumination such that the *I-B* curve stretched over the zero of current (so $I_P$ is negative while $I_{AP}$ is positive), the changes in current flow direction depended on the magnetic field (Fig. 3A). For a light intensity 4.67 mW/cm$^2$, we obtained $I_P = -I_{AP}$ and the MSP device acted as a perfect magnetic current inverter (Fig. 3B). Given the large changes in the output current obtained for different light illumination, together with the constant $\Delta I$ current value between parallel and antiparallel magnetic configurations (see Fig. 3, A and C), the traditional calculation of magnetocurrent does not have a clear meaning in this scenario. For example, and as depicted in Fig. 3C, we can achieve a *MC* tending to infinite for particular light intensity irradiation values, stressing again the link between spin transport and light. This very large figure-of-merit could be useful in complex magnetic sensors, especially in those requiring a negligible baseline, or as a light sensor. Note that the specific value of the light intensity at which the MC maximizes depends on the voltage applied, and hence an extra tunability could be obtained.

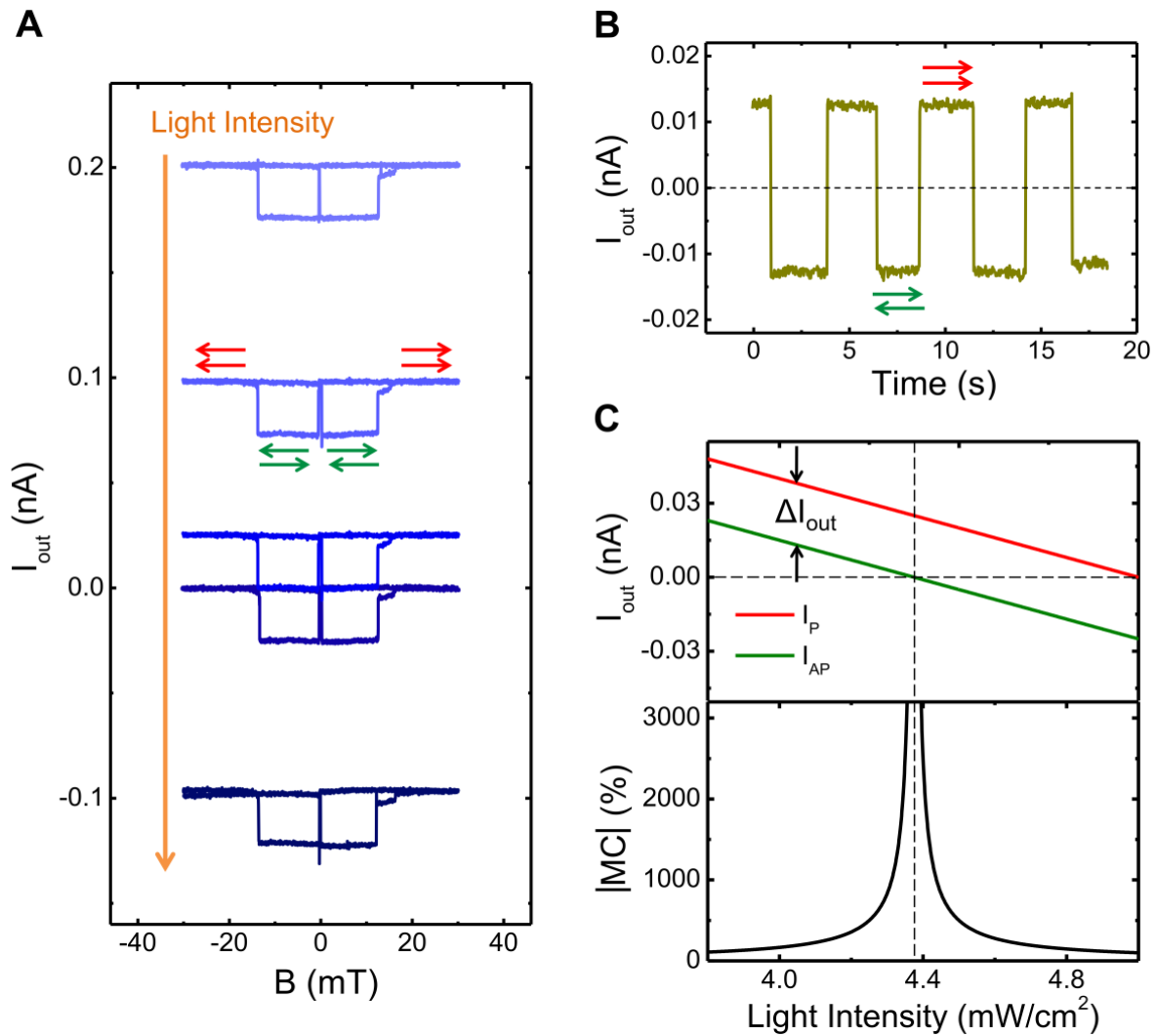

**Fig. 3. Operation of the device at a constant applied voltage bias under different light irradiation conditions.** (**A**) Manipulation of *I-B* curves in a $C_{60}$ spin photovoltaic device by the irradiated light intensity under a constant bias of 10 mV at 80 K. (**B**) In the particular case of $I_{AP} = -I_P$, the flow direction of the current can be freely inverted by the magnetic field controlling the orientation of the magnetic electrodes without changes in its absolute value. (**C**) Output current $I_{out}$ ($I_{AP}$ and $I_P$) and normalized magnetocurrent response versus the irradiated light intensity; the curves have been calculated from discrete data measured as exemplified in panel (A).

Because of the intermixing between different degrees of freedom, the manipulation of the *I-B* curves could also be achieved by modifying the applied bias under a constant light irradiation. In this case, both the magnitude and the sign of the *I-B* curves were modified (see Fig. 4A). The comparison between both modes of operation, light intensity and applied bias, were observed directly in Fig. 4B.

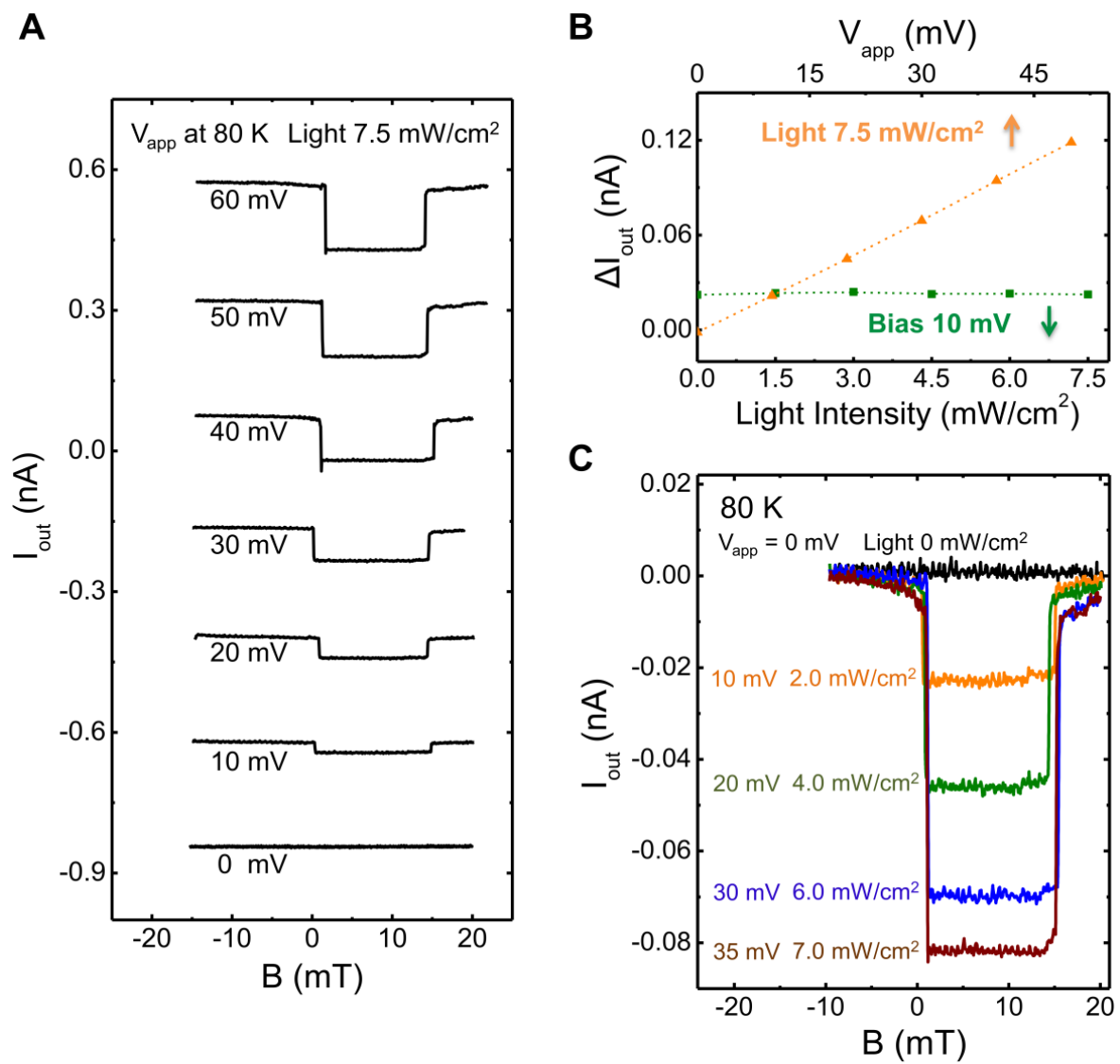

**Fig. 4. Operation of the device at a constant light irradiation condition under different applied voltage biases, together with the more complex electro-optical modulation with varying voltage and light.** (**A**) Manipulation of *I-B* curves in a $C_{60}$ spin photovoltaic device under different applied biases and constant light irradiation of 7.5 mW/cm$^2$ at 80 K. (**B**) Change in current under the application of a magnetic field ($\Delta I$) obtained as a function of both applied light irradiation and applied voltage bias ($V_{app}$). (**C**) Electro-optical modulation, with varying both the applied voltage bias and the light irradiation. The input values are selected in such a way that $V_{app}$ exactly cancels $V_{OC,P}$. The device output is a spin photo-generated current.

The qualitative difference between the light and bias manipulation of the magnetocurrent displayed in Fig. 4B was shown schematically in Fig. 2. The change in the applied bias modifies the amount of accumulated spin charge carriers at the anode/molecule interface, while the change in light irradiation modifies the amount of photogenerated carriers. If we adjusted in each case the applied bias to cancel the open circuit bias created by the light, a spin photo-generated current can be modulated (Fig. 4C and S4). In this case, the output current of the device was set to zero for a parallel orientation of the magnetization of the electrodes and the change in the magnetization alignment led to the generation of the spin-polarized output current (equal to $\Delta I$). Moreover, the magnetocurrent recorded is a pure spin-related signal was caused entirely by the spin valve effect. This effect could lead to a more accurate fitting of the experimental data to some existing theoretical models and, in general, to a better understanding of the spin transport in molecular materials (*34, 35*). Moreover, and with potential relevance to applications of molecular opto-electronics, the magneto-photocurrent signal has also been observed at room temperature (see Fig. S3). In traditional molecular spintronics devices, such as spin valves, there is a relatively large power consumption created by the background current. However, since in our case the base current is set to zero due to the electro-optical modulation, the operating power $P$ ($= I^2R$) for obtaining the same magneto-current response is was reduced < 1% when compared to standard molecular spin valves (*24, 26, 27*).

**Acknowledgments:**

The authors acknowledge financial support from the Ministry of Science and Technology of China (2016YFA0200700, 2017YFA0206600); the European Union 7th Framework Programme under the Marie Curie Actions (256470-ITAMOSCINOM), the European Research Council (257654-SPINTROS) and the NMP project (263104-HINTS); the Spanish

# Supplementary Materials for

## A molecular-spin photovoltaic device

Xiangnan Sun[1,2], Saül Vélez[3], Ainhoa Atxabal[3], Amilcar Bedoya-Pinto[4], Subir Parui[3], Xiangwei Zhu[1,2], Roger Llopis[3], Fèlix Casanova[3,5], Luis E. Hueso[3,5]*

correspondence to: l.hueso@nanogune.eu

**This PDF file includes:**

Materials and Methods
Figs. S1 to S4
References

**Materials and Methods**

Co/AlO$_x$/C$_{60}$/Ni$_{80}$Fe$_{20}$ vertical devices have been fabricated in-situ in a UHV dual chamber evaporator (base pressure < 10$^{-9}$ mbar). The junction areas range from 150 × 200 to 500 × 200 μm$^2$. In all samples, eight 11-nm-thick Co lines are deposited as bottom electrodes. Then, a 1.5-nm-thick Al layer was deposited on top and was semi-oxidized in-situ by oxygen plasma, which named as leaky AlO$_x$ barrier. This barrier is employed to isolate the active ferromagnetic (FM) Co from the C$_{60}$ layers in the vertical-geometry device, and also demonstrated to be crucial for the (spin) carrier injection (*36-38*). The 15 nm-thick C$_{60}$ layer has been evaporated from a Knudsen cell at a rate of 0.1 Å/s and deposited onto the AlO$_x$ though a shadow mask. Finally, an 11-nm-thick Ni$_{80}$Fe$_{20}$ line has been deposited in cross-bar alignment to the top electrode. Liquid nitrogen has been used to cool down the sample during the deposition of metals and molecules to assure the quality of the MSP device (*36-38*).

The metals (Co, Al and Ni$_{80}$Fe$_{20}$) were purchased from Lesker (99.95% purity), and have been deposited by e-beam evaporation in one of the chambers with a deposition rate of ~1 Å/s (the first 2 nm of Ni$_{80}$Fe$_{20}$ has been deposited with a lower rate of ~0.1 Å/s for a gentle deposition of the metal atoms onto the organic layer). The C$_{60}$ was purchased from Aldrich (sublimed grade, 99.9%) and used as received. The thickness of the molecular layers is controlled by the crystal monitor in our deposition system during the deposition, and subsequently measured by both Atomic force microscope (AFM) and X-Ray Reflectivity (XRR).

Electrical characterization was performed in vacuum (or in air atmosphere) with a magnetic-field-equipped Lakeshore probe station with variable temperature, from 6 K to 350 K. A Keithley 4200 semiconductor analysis system has been used to record current-voltage (I-V) and magnetoconductance (MC) curves with a 4-probe method.

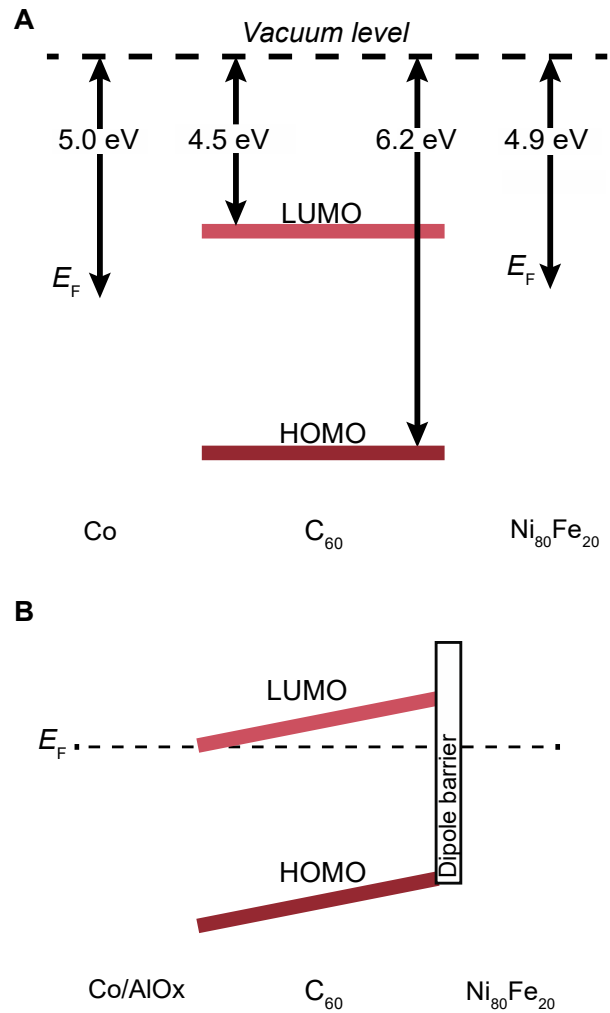

**Fig. S1.**
**Rigid band energy diagram of the C$_{60}$ based device.**
(A) Rigid band energy diagram of the individual layers of the device. The Fermi energy ($E_F$) of the metals, together with the lowest unoccupied molecular orbital (LUMO) and highest occupied molecular orbital (HOMO) energies of C$_{60}$, are indicated. (B) Interface formation at thermal equilibrium for unbiased conditions (short-circuit mode) in which the Fermi levels of the metals are aligned. Considering the experimental optoelectronic response of our device, we infer that the presence of the leaky AlO$_x$ layer minimizes the Co/C$_{60}$ LUMO energy barrier. In similar terms, on the other side, the energy barrier between the C$_{60}$ and Ni$_{80}$Fe$_{20}$ is indicated with a possible dipole barrier in between (*39*).

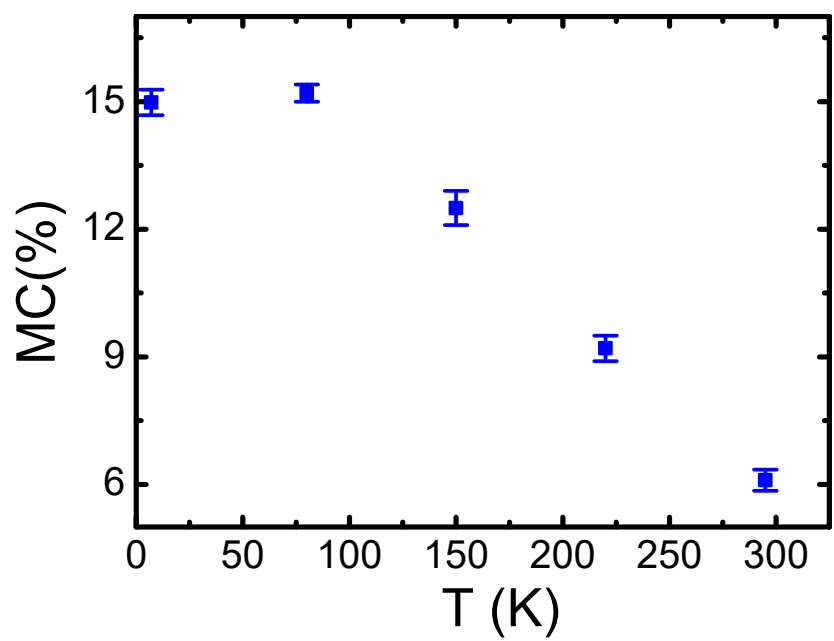

**Fig. S2**
Magnetocurrent (MC) values as a function of temperature.

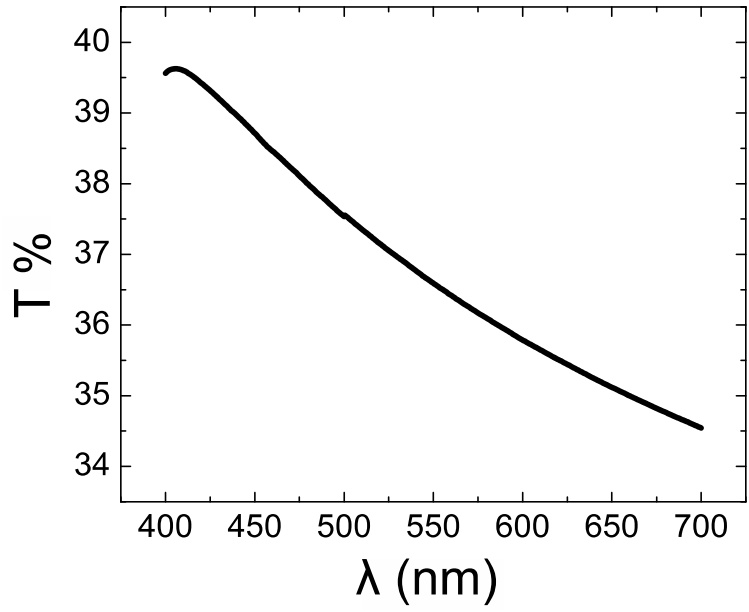

**Fig. S3**
Transmittance of a 11 nm-thick $Ni_{80}Fe_{20}$ film as a function of the incident wavelenght.

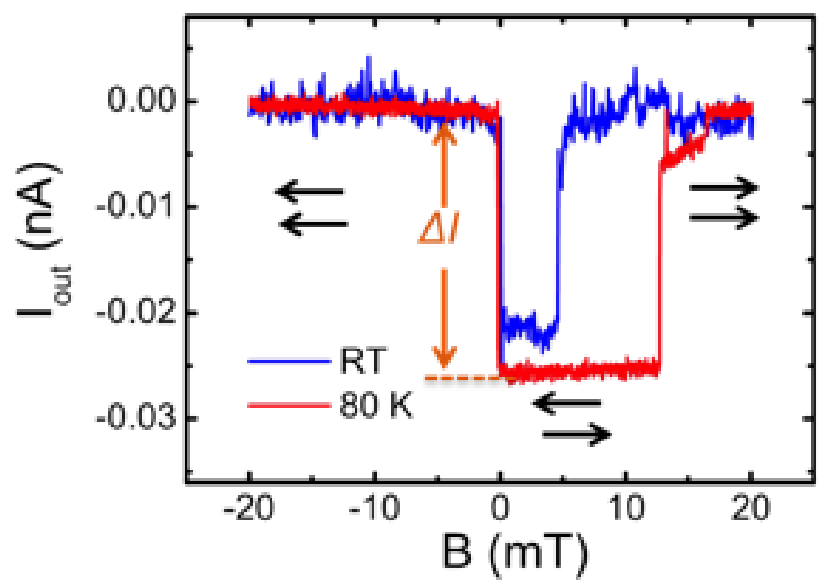

**Figure S4.**
Pure photocurrent response to the magnetic field at RT and 80K.